\documentclass[12pt,a4paper]{article}
\usepackage{graphics}
\usepackage{amsmath}
\usepackage{amsfonts}
\input{psfig}
\newcommand{\newsec}[1]{\section{#1}}
 
\newcommand{\bpm}{\left( \begin{matrix}}
\newcommand{\epm}{\end{matrix} \right)}
\newcommand{\ba}{\begin{eqnarray}} \newcommand{\ea}{\end{eqnarray}}

\newcommand{\ap}{\alpha '} 
\newcommand{\ra}{\rightarrow} 
\newcommand{\Om}{\Omega}
\newcommand{\D}{\partial}

\newcommand{\N}{{\cal N}}
\newcommand{\ep} {\varepsilon}
\newcommand{\la}{\theta} 
\begin{document}

\begin{titlepage}
\rightline{hep-th/0008141}
\rightline{TAUP-2633-00}
\vskip 1cm
\centerline{{\Large \bf Supergravity and Worldvolume Physics}} 
\vspace{6pt}
\centerline{{\Large \bf in the Dual Description of $\N=1^*$ Theory}}
\vskip 1cm
\centerline{ Y. Kinar, A. Loewy, E. Schreiber, J. Sonnenschein, S. Yankielowicz}
\vskip 1cm
\begin{center}
\em School of Physics and Astronomy
\\Beverly and Raymond Sackler Faculty of Exact Sciences
\\Tel Aviv University, Ramat Aviv, 69978, Israel
\\ \{yaronki, loewy2, schreib, cobi, shimonya\}{\bf @}post.tau.ac.il
\end{center}
\vskip 1cm
We look at the string theory dual of the $\N=1^*$ theory, involving
5--branes, which was
recently proposed by Polchinski and Strassler \cite{PolStr}. We argue
that SUGRA alone is not enough in order to obtain the correct
screening and confinement behaviour of the various massive field theory vacua,
but that appropriate worldvolume phenomena of the 5--branes must be included.
We therefore work within the SUGRA approximation, also taking into account 
the brane dynamics,
and classify all the SUGRA configurations.
In this level of analysis, 
we find multiple valid 
configurations for every given vacuum. 
We discuss some possible resolutions of this perplexing result.
We also consider the spectrum of asymptotic states, and discuss the global
symmetries of the SUGRA solution of the $\N=1^*$ theory and of
the $\N=0^*$ theory obtained from it by explicit supersymmetry breaking. 
\end{titlepage}

\newsec{Introduction}

The AdS/CFT correspondence \cite{Mald,witten,gubser} was originally 
formulated for theories with 16 supersymmetries. A string theory dual 
of a non--conformal confining theory with 4 supersymmetries was
recently suggested by Polchinski and Strassler \cite{PolStr}. 
This theory has come to be known as the $\N=1^*$ theory. 
It is derived from the $\N=4$ theory by adding mass terms to the three
$\N=1$ chiral multiplets. 

The supergravity dual suggested in \cite{PolStr} is obtained by
turning on the dual supergravity modes of the mass terms added. 
This results in the creation
of a 5--brane wrapped on an $S^2$. A five dimensional gauged SUGRA
dual of the same theory was presented in \cite{girardello}. 
Although the supergravity equations
were only solved to first order in the perturbation, certain
qualitative features emerge. The most striking feature that can,
according to \cite{PolStr}, be obtained from the supergravity dual 
is the existence of confinement in some of the vacua of this theory. Other related works that followed are \cite{peet,freedman,frey,bachas,pilch,aharony1,aharony2,bena}.

The purpose of this paper is to point out some subtleties in the 
duality between the $\N=1^*$ theory and the supergravity 
description suggested in \cite{PolStr}. 
In section \ref{sec:review} we review the main features of the $\N=1^*$ theory 
from the field--theoretic point of view as well as from the one of the 
SUGRA dual solution.

In section \ref{sec:screening} we address the inadequacy of SUGRA
to describe the screening of charges in the dual field theory.
It was argued in \cite{PolStr} that using the SUGRA duals
of the various vacua of the field theory,
one can manifest screening, or the existence of confinement with the
appropriate string tension.
The configuration of $p$ D5 branes
corresponds to the unbroken $SU(p)$ vacuum of $SU(N)$ (with $p|N$). 
We show that except for the special case $p = 1$ (a Higgs vacuum) 
discussed in \cite{PolStr}, one needs to incorporate 
the worldvolume theory on the 5--branes in order to get the right 
set of confined and screened charges. A purely SUGRA analysis leads to
the wrong conclusion of total screening in {\em all} of those vacua.
Analogously, a purely SUGRA analysis of $q$ NS5 branes 
configuration (with $q|N$) gives an erroneous result of monopole 
screening instead of the correct $SU(N/q)$ vacuum.

Polchinski and Strassler 
have argued that within their approximation there is only one valid
SUGRA description for each field theory vacuum. 
In section \ref{sec:classification} we find and describe the
multiplicity of SUGRA descriptions of each field theory vacuum.
We argue that, in this level of analysis, there is a multitude of 
valid descriptions for each such vacuum. 
We then discuss some possible resolutions of this perplexing result. 
Some details are given in the appendix.

In section \ref{sec:glueball} we consider the asymptotic states in the
theory.
We do not calculate the exact glueball spectrum, but work with the
simplified analogue of a minimally coupled scalar. 
This is intended to give a rough idea of the glueballs mass scale. 
Section \ref{sec:symmetry} deals with the global symmetries of the
theory and the possibilities for  complete breaking of supersymmetry.
Finally, section \ref{sec:summary} is a summary and discussion of our results.

\newsec{A Review of $\N=1^*$ Theory}
\label{sec:review}
   
The starting point is the $\N=4$ SYM theory, whose field content is --
in the $\N=1$ language --  three chiral superfields $\Phi_i, \;\;(
i=1,2,3)$, and a vector superfield, all in the adjoint representation of
the gauge group. The superpotential is
\begin{equation} 
\label{supot}
W = \frac{\sqrt{2}}{3 g^2_{YM}} tr \left( \epsilon^{ijk} \Phi_i \left[
    \Phi_j , \Phi_k \right] \right).
\end{equation}
The theory possesses $SO(6) \simeq SU(4)$ R--symmetry of which $SU(3)$
(rotating the three chiral superfields) and $U(1)$ (the $\N=1$
R-symmetry) are manifest in the superpotential. The deformations
which were considered are the turning on of mass terms -- 
either a mass $m_4$
for the gaugino, in which case supersymmetry is completely broken, or
masses $m_1, m_2, m_3$ for the chiral superfields, 
keeping $\N=1$ supersymmetry. The latter deformation
is performed by adding to the superpotential 
\begin{equation}
\label{pert}
\Delta W = \frac{1}{g_{YM}^2} \left(m_1 tr \Phi_1^2 + m_2 tr \Phi_2^2
  + m_3 tr \Phi_3^2 \right),
\end{equation}
thus breaking the R--symmetry down to $U(1)$ (which is further broken by
anomaly). An interesting and simpler case is the $SO(3)$ invariant
theory where all masses are chosen to be equal.
Ideally, one would like to take $m_i$ to infinity, thus decoupling the
chiral superfields and ending up with a pure $\N=1$ SYM. However, this
cannot be done within the AdS/CFT correspondence since the dynamically
generated energy scale is 
\begin{equation} 
\label{LambdaQCD}
\Lambda^3 = m_1 m_2 m_3 \exp{(2 \pi i \tau / N)} \sim m_1 m_2 m_3 \exp{(-
  8 \pi^2/\lambda)}
\end{equation}
(where $\tau = \frac{\theta}{2 \pi} + \frac{4 \pi i}{g_{YM}^2}$, and 
$\lambda = g_{YM}^2 N = 4 \pi g N$ is the 't Hooft coupling constant),
 and decoupling can
only be achieved with $\lambda \rightarrow 0$ which invalidates the
SUGRA approximation. One has, therefore, to deal with the
$\N=1^*$ theory in which the fermions and scalars are not decoupled. 
Considering the string theory dual of this perturbation, the fermion mass
matrix transforms as the $\bf{\overline{10}}$ of $SU(4)$ and was shown
\cite{PolStr} to correspond to the lowest $S^5$ spherical harmonic,
non-normalizable mode of the 3-form $G_3 = F_3 - \tau H_3$ on 
$AdS_5 \times S^5$. 
The asymptotic behavior of this form should correspond to a tensor that
transforms appropriately under $SO(6)$. Such a tensor can be formed by
defining complex coordinates 
\begin{equation}
z_i \equiv \frac{x_{i+3} + i x_{i+6}}{\sqrt{2}} 
    \equiv \frac{w_i + i y_i}{\sqrt{2}}
\end{equation}
and regarding the form 
\begin{equation}
\label{T3}
T_3 = m_1 dz_1 \wedge d \bar{z}_2 \wedge d \bar{z}_3 + m_2 d \bar{z}_1
\wedge dz_2 \wedge d \bar{z}_3 + m_3 d \bar{z}_1 \wedge d \bar{z}_2
\wedge dz_3 + m_4 dz_1 \wedge dz_2 \wedge dz_3
\end{equation}
and descendents thereof, where the $x_i$'s are the six  coordinates 
transverse to the D3 branes.
 
\subsection{Field Theory Vacua}

The $\N = 1^*$ theory has been investigated by Vafa and Witten who have
found the classical vacua \cite{VafWit}, and further by Donagi and
Witten who have considered the quantum vacua \cite{donagi}.
A supersymmetric vacuum must have zero F- and D- terms.
Assuming the three masses $m_1,m_2,m_3$ are nonzero, 
the chiral superfields may be
rescaled such that all masses are equal to $m$, and one finds
\ba
F_i &=& \frac{m}{\sqrt{2}} \phi_i + \epsilon_{ijk} [\phi_j,\phi_k] \\
D   &=& \sum_{i=1}^3 [\phi_i^\dagger,\phi_i]
\ea
where the $\phi$'s are the lowest (scalar) components of the chiral
superfields. The F-term equations dictate that a vacuum is determined
by an $N$ dimensional representation of $SU(2)$, i.e., a homomorphism
from the algebra of $SU(2)$ to the one of the gauge group
$SU(N)$. These vacua are to be modded out by the gauge 
transformations and imposed with the D-term condition. This amounts to 
modding by the complexified gauge group $SL(N,\bf{C})$, or, alternatively, 
considering the conjugacy classes of the homomorphisms. As $SU(2)$
has, up to conjugacy, 
a single irreducible representation of every dimension, we
end up with a discrete set of vacua (corresponding to the choice of
the {\it reducible} representations we make). The apparent $SO(3)$  
global symmetry is equivalent to a gauge transformation of an 
$SO(3) \subset SU(N)$.
This can also be shown explicitly -- consider an
infinitesimal $SO(3)$ transformation, e.g. a rotation in the $1-2$ plane 
\begin{equation}
\label{infiso3}
\delta \left( \begin{array}{c} \phi_1 \\ \phi_2 \\ \phi_3 \end{array}
\right) = \left( \begin{array}{c} \ep \phi_2 \\ - \ep \phi_1 \\ 0
  \end{array} \right). 
\end{equation}
This transformation can also be performed (around a classical vacuum)
by commuting with $\phi_3$, which by the D-term condition may be taken
as Hermitian, thus being a generator of $SU(N)$. This can also be done
for $\phi_1$ and $\phi_2$, showing that the global $SO(3)$
mixes with the gauge group. 
Therefore, in each of the (non confining) vacua there is a new $SO(3)$, 
which is a combination of the original $SO(3)$ and the gauge group. 
This new global symmetry will presumably be the isometry of the 
SUGRA description of that vacuum.

To conclude, in the classical analysis, each of the discrete set of vacua
of the $\N = 1^*$ theory is described by assigning, as {\em vev}s of
the scalars, an $N$-dimensional,
generically reducible, representation of $SU(2)$. Using
the above mentioned uniqueness of the {\it irreducible} representations,
each vacuum corresponds to a partition of $N$ into positive
integers. The resulting unbroken gauge group of the theory depends
upon this choice, varying from the full $SU(N)$ confining phase
(corresponding to $N$ copies of the trivial representation, 
i.e. $\phi_i = 0, \;\; i=1,2,3$) to the
completely Higgsed phase (The irreducible $N$-dimensional
representation). The massive vacua, those with a mass gap, correspond
to divisors of $N$, $N = p \cdot q$, where the unbroken gauge group $SU(p)$
rotates the $p$ copies of the $q$-dimensional representation. 
A partition $N = \sum_{d=1}^N d \, k_d$ with more than one nonzero
$k_d$ corresponds to a
Coulomb vacuum where the gauge group includes a $U(1)$ factor and
there is no mass gap.

Classification of the quantum vacua is more complex. Considering only
the massive vacua, the gaugino condensate 
$\langle \lambda_4 \lambda_4 \rangle$
splits each $SU(p)$ vacuum into $p$ different quantum
vacua. Therefore, the number of quantum massive vacua of the $SU(N)$ 
$\N = 1^*$ theory is given by the sum of the divisors of $N$, which
we shall denote by $\sigma_1(N) \equiv \sum_{p|N} p$.  
These vacua are in a one-to-one correspondence with
't Hooft classification \cite{tHo} : consider the 
lattice $L \equiv \bf{Z}_N \times \bf{Z}_N$ of external 
electric and magnetic charges (one
factor corresponds to the center of the gauge group while the other to
the fundamental group $\pi_1 [SU(N)/\bf{Z}_N]$).
Each vacuum is characterized by a subgroup $P$ of the
condensed charges. The massive vacua correspond to subgroups
of order $N$, and in these vacua the screened charges are exactly the
condensed ones. In the confining vacuum, the
magnetic monopoles are condensed and the subgroup is generated by
$(0,1)$, while in the Higgs
vacuum the condensed charges are the electric ones, and the subgroup 
is generated by $(1,0)$. In general, an $SU(p)$ vacuum
corresponds to a subgroup generated by $(p,0)$ and $(s,q)$, with
$s=0 \dots p-1$ distinguishing between confinement and various oblique
confinements. This classification exhausts the order $N$ subgroups,
and also clarifies the action of the
$SL(2,\bf{Z})$ duality which is inherited from the Montonen--Olive
symmetry in $\N=4$ SYM. The elements of the duality group 
$SL(2,\bf{Z}_N)$ act in the natural way on elements of $L$, permuting
between the various massive vacua. 

The complete superpotential of $\N=1^*$ theory, including quantum
corrections, has been computed \cite{dorey1}, and it was shown that
the aforementioned enumeration of vacua still holds.

We can give a simple physical picture of confinement.
With $N = p \cdot q$, the $adj$ representation of $U(N)$ decomposes
to $(adj , adj)$ in the obvious $U(p) \times U(q) \subset U(N)$.
For $SU(N)$ the $adj$ decomposes into 
$(adj , adj) \oplus (1 , adj) \oplus (adj , 1)$ of 
$SU(p) \times SU(q) \subset SU(N)$.
When we further break the $SU(q)$ factor, it decomposes to $q^2$
copies of the $SU(p)$ $adj$ and to singlets.
For the massive vacua the gluons of the unperturbed theory
do not produce fundamental representations. It follows that external
fundamental charges (``massive quarks'') cannot be screened by pair
creation from the vacuum. In conclusion, for $1 < p,\; p | N$, we
indeed expect confinement of basic electric charges in the $SU(p)$ vacuum.

\subsection{String Dual Description}

As mentioned before, the mass deformation corresponds to turning on
the SUGRA form $G_3$. According to Myers' dielectric brane mechanism
\cite{Myers}, this results in an induced 5--brane dipole, or more
specifically, considering the equal masses case, a 5--brane which wraps
an $S^2$ equator of the
unperturbed $S^5$ at a constant AdS radius and also spans 
the four dimensional Minkowski space. This 5--brane has a 
D3 charge distribution on it due to the original D3 branes. 
Different massive vacua correspond to different kinds of
wrapped 5--branes. The location (in the AdS radius) of these branes is
determined by the minimum of the brane action, which was calculated by
Polchinski and Strassler \cite{PolStr} to first order in a small parameter 
$g |M|^2 / N$ (where $M = c \tau + d$ for a $(c,d)$ 5--brane). The complete
metric and dilaton background of these vacua were also derived, as well
as the solution of $G_3$. 

The $SU(p)$ vacuum, with $N=p \cdot q$, was described by two alternative dual
configurations -- either $p$ coincident D5 branes or $q$ coincident
NS5 branes. The first description seems plausible, 
because of the low energy theory of open strings ending on the D5 branes, 
but the second description is more  mysterious. The complete metric
for the $p$ D5 solution is
\begin{equation}
\label{metric}
ds^2 = Z^{-1/2} dx_{0123}^2 +
       Z^{1/2} (dy^2 + y^2 d\Omega_y^2 + dw^2) +
       Z^{1/2} \Omega^{1/2} w^2 d\Omega_w^2
\end{equation}
(corresponding to a brane located at $|\vec{w}|=r_0$ and $|\vec{y}|=0$
where $\vec{w},\vec{y}$ are both three dimensional) with
\[
Z \equiv \frac{R^4}{\rho_+^2 \rho_-^2}, \;\;
\ \ \ \ \ \ \Omega = \left[\frac{\rho_-^2}{\rho_-^2 + \rho_c^2} \right]^2
\]
where, for 't Hooft coupling $\lambda$ 
of the unperturbed theory,
\[
R^4/{\alpha'}^2 = \lambda, \;\; 
\rho_\pm = (y^2 + [w \pm r_0]^2)^{1/2}, \;\;
\rho_c = \frac{{\alpha'}^2 R^2 m}{2}, \;\; r_0 = \pi \alpha' m q.
\]
The nontrivial dilaton is
\[
e^{2 \phi} = g^2 \frac{\rho_-^2}{\rho_-^2 + \rho_c^2}.
\]
The conditions for this solution to be valid are: a small string
coupling constant, small curvature (i.e. large radius of the
transverse sphere) and a small perturbation around the solution with
$m = 0$ (the small parameter mentioned above). For the two
descriptions of $SU(p)$ with zero $\theta$-angle, the conditions are 
\begin{table*}[htbp]
  \begin{center}
\begin{tabular}{|l||c|c|}
\hline
                   & $p$ D5's    & $q$ NS5's   \\
\hline 
\hline 
small perturbation      & $q \gg g p$ & $q \ll g p$ \\
\hline
small coupling          & $g \ll 1$   & $g \ll 1$   \\
\hline
large transverse sphere & $g p \gg 1$ & $q \gg 1$   \\
\hline 
\end{tabular}
\end{center}
\caption{Validity conditions of the solutions for the SU($p$) vacuum}
\end{table*}

The ranges of validity of these two descriptions have no
overlap, but for each large enough value of $\lambda$ there are valid
solutions considering both descriptions.

\newsec{Screening and Confinement via Stringy Effects}
\label{sec:screening}

Polchinski and Strassler look at the behavior of Wilson loops for the
fully Higgsed vacuum, using the single D5 brane vacuum, although the
SUGRA solution does not obey the aforementioned conditions. They find
complete screening, as they expect. We shall now claim that the SUGRA
description is inadequate even for the ``electric'' case (fundamental
strings ending on D5 branes), and that stringy effects of the brane
world-volume theory must be taken into account.

\subsection{Wilson--'t Hooft Loops as String Probes Approaching 5--branes} 

Let us first briefly recall the case of flat D5 branes. The metric is
given by 
\begin{equation}
\label{D5metric}
ds^2 = \frac{\rho}{R'} dx_{012345}^2 + \frac{R'}{\rho} d\rho^2 + 
R' \rho \, d\Omega_3^2.
\end{equation}
An open string probe in this background, with both ends fixed on the
Minkowski boundary of (\ref{D5metric}), 
gives the dual description of
the Wilson loop of a quark anti--quark pair \cite{Mal2,ReYe}.
A simple scaling argument \cite{BISY2} shows that 
a smooth configuration obeying the equations of
motion must have the quarks at a fixed distance apart (this distance is a
numerical multiple of $R'$). For smaller
distances, the string will be pulled to the boundary 
($\rho \rightarrow \infty$)
and there is no stable solution. For larger distances only the 
non smooth configuration of two ``bare'' quarks, represented by two
straight ``vertical'' strings from the boundary to the origin, is allowed. 
This configuration corresponds to a {\em screening} behavior.
 This correspondence can be understood in the following manner. 
Consider a 
  ``horizontal'' string segment connecting the two quark lines at some
 given value of $\rho =\rho_0$. This segment has a Nambu--Goto action 
which is linear
with the separation distance and with a ``string tension'' 
 given in the SUGRA approximation as 
\begin{equation}
\sigma(\rho_0) = {\alpha'}^{-1} \sqrt{G_{tt} G_{xx}} 
= \frac{\rho_0}{\alpha' R'}
\end{equation}
If the SUGRA picture is valid all the way down to  $\rho \rightarrow 0$
it is clear that one finds a zero string tension, namely, screening. 
However, this SUGRA analysis cannot give the complete answer.  
For $\rho \sim \alpha'/R'$, the curvature becomes large in string units, 
and the SUGRA solution and description of the Wilson loop  
ceases to be valid. 
 The smallest
reliable tension SUGRA can give us is
$\sigma_0 = \sigma(\alpha'/R') \sim 1/{R'}^2$.
The question what is the tension
when the string probe reaches smaller values of $\rho$, and whether 
it vanishes for $\rho \rightarrow 0$, 
cannot be answered by SUGRA alone.

This conclusion holds for any number of flat D5 branes, as their
number only enters through $R'$.

We now turn to the actual metric (\ref{metric}) 
of the  5--branes wrapped on $S^2$. 
Following \cite{PolStr}, we analyze the Nambu--Goto action, ignoring
the contribution of the NS potential.   
Very far from the sphere, the D3
charge dominates and the metric is asymptotically
$AdS_5 \times S^5$. This, of course, allows for small distance 
Wilson loops, in contrary to the flat D5 case. Near the sphere,
however, the metric resembles that of a stack of flat D5 branes. 
Indeed, as an expansion in small $\rho_-$ shows, $\rho_-$ plays the
role of $\rho$ in (\ref{D5metric}), the surface of the sphere seems
flat, and the two coordinates of $d\Omega_w$, properly scaled, join
the four dimensional Minkowski space to form a six dimensional one, while
the extra angle between $y$ and $w-r_0$ joins the two coordinates of 
$d\Omega_y$ to form $d\Omega_3$.
We also have $R' = R^2/r_0$.
We therefore expect that in the naive SUGRA analysis 
there is a maximal value for the distance $L$ between the ends of a 
smooth Wilson loop. As the quarks are
moved farther apart, the string touches the stack of branes and the
favored configuration becomes necessarily that of two bare quarks,
as in the flat D5 case.
The exact shape of the Wilson loop depends on the boundary condition
for its position in the $w , y$ plane, but at least for the case of
the loops with $y = 0 ,\, r_0 \le w \le \infty$, this is 
proven by the theorems in \cite{KSS2}. The analysis is also performed
for a similar system in \cite{BraSfe}. Therefore, the proposed SUGRA
picture shows {\em complete screening}, even for the $SU(p)$ vacuum, where
we expect {\em confinement} of the basic electric charges.
Hence,  the
correct description necessitates stringy effects and  we
definitely cannot rely on SUGRA to discern between confinement and screening!

However, The minimum tension for a bit of string, artificially 
placed ``horizontally'' in a region where SUGRA {\em is} reliable, 
gives an upper bound for the actual string tension. This provides us
with a consistency check of the proposed duality. 
The minimum reliable SUGRA tension takes the value 
$\sigma_0 \sim r_0^2/R^4 \propto m^2 \cdot q / g p$.
We remember that the necessary condition for the metric to be valid as a
first order approximation, i.e. for the perturbation to be small,
is $q \gg g p$. We therefore have $\sigma_0 \gg m^2$. On the other
hand, all the three consistency conditions listed above imply 
$\lambda \gg 1$ and therefore, by (\ref{LambdaQCD}), 
$\Lambda_{YM} \sim m \exp (-8 \pi^2/3\lambda) \approx m$. We expect the
string tension, if confinement occurs as in the pure $\N=1$ case, 
to be of order $\Lambda_{YM}^2 \approx m^2$. 
We indeed find that the upper bound $\sigma_0$ is higher than that.

The dual description of the $SU(p)$ vacuum, using NS5 branes, can be seen,
in an analogous way to the D5 description, to give complete monopole 
screening from the SUGRA analysis.
Again, only stringy effects can lead to the correct $SU(p)$ vacuum 
confinement of basic magnetic monopoles.  

\subsection{Strings Ending on Branes and ``Electric'' Screening}

The field--theoretic picture of the $SU(p)$ vacuum of ${\cal N} = 1^*$ 
demands that there is confinement of basic electric charges. We have
shown that SUGRA alone cannot account for this phenomenon. 
A full understanding of string theory in  regions of spacetime which
include branes, where the curvature is high and there are other
strong SUGRA fields, should 
provide the correct answer. Until such an understanding is
available, we should assume some effects associated with
 the 5--branes worldvolume physics. 

In the picture of $p$ wrapped D5 branes of the $SU(p)$ vacuum, a 
$(1 , 0)$ string
(a fundamental string) ending on these branes is a quark of the
$SU(p)$ worldvolume theory on the $p$ D5 branes.  A non Abelian
flux tube inside the branes should confine a quark anti--quark pair
into a singlet (a meson).    
Such a singlet is trivial from the point of view of the
$SU(N)$ theory. However,  $p$ fundamental
strings form a singlet of the $SU(p)$ worldvolume theory 
(an ``electric'' baryon), and this corresponds to 
 a non trivial screening of $(p , 0)$
charges in the $SU(N)$ theory, and the breaking
of the   $SU(N)$ group down to  $SU(p)$.

This picture can be generalized using the $SL(2,\bf{Z})$ symmetry of the
IIB theory. On a wrapped stack of $(c , d) = p' \cdot (c' , d')$
5--branes, where $p' | N$ and $c',d'$ are mutually prime, a 
$(d' , c')$ string may naively end (in our conventions), 
but only a bunch of $p'$ such
strings can indeed do so. For $q$ NS5 branes (with $N = p \cdot q$), 
this means screening of
$q$ D1 strings, or of $q$ monopoles, as required in the non oblique
$SU(p)$ vacuum.

However, there is a subtlety here. Suppose there is a wrapped 
$(c , d)$ brane with $f = \gcd(c,d),\;\; p' = \gcd(f,N)$. Usually such a
configuration is viewed as $f$ coincident branes, but we claim that
in our scenario it should be viewed as $p'$ coincident branes. Indeed,
the dielectric effect gives the branes a common radius $r_0$ at equilibrium. 
$r_0$ is proportional to the ratio of D3 brane to 5--brane
charges. For any radius different from $r_0$, there is a positive
energy density of the configuration and supersymmetry is completely broken. 
However, the D3 charge, $N$, embedded in these 5--branes cannot 
be equally divided between $f$ branes. There is an energy barrier
between the one--shell configuration and any configuration of
$f$ displaced branes (a stable such configuration, where the branes
form several separated shells, necessarily corresponds to a 
Coulomb vacuum). The energy barrier per volume is proportional to $N$,
so cannot be overlooked in the $N \rightarrow \infty$ limit, and is
infinite for the infinite volume limit of the field theory. 
We conclude that a $(c' , d')$  5--brane
is to be considered a single brane, if its D3 charge is mutually prime
with $\gcd(c' , d')$, even if the latter is bigger than one. 
For ``electric'' screening this means
that the $(c , d)$ 5--brane allows screening of $p'$ copies of 
$(d'' , c'') \equiv (d/f , c/f)$  
strings (or charges).
Note that this result is also true modulu $N$.

Such configurations, in which the D3 charge cannot be evenly split
between the (naively) multiple 5--branes, may seem superfluous.
They are present in the SUGRA analysis, but, of course, in the full
string--theoretic treatment they might either disappear or display different
dynamics. We return to this point at the end of the next section. 

\subsection{Baryon Vertices and ``Magnetic'' Screening}

Polchinski and Strassler describe D3 branes filling the $S^2$ sphere on
which the 5--branes are wrapped, which behave as baryon vertices. 
These baryon vertices arise through the Hanany--Witten effect
\cite{HanWit}, when a baryon vertex of the unperturbed ${\cal N} = 4$
theory, which is a 5--brane wrapping the $S^5$, contracts and moves
through the 5--brane $S^2$ shell.
For the $SU(p)$ vacuum described by $p$ wrapped D5 branes, each D5
brane has a dissolved D3 charge of $q$, and its
analysis shows that the junction of a D3 ball with a D5 wrapped on $S^2$
must support strings with total $D1$ charge of $q$. Polchinski and
Strassler do not constrain the $F1$ charge of the strings, but simply
take it as zero. This mechanism explains the ``magnetic'' screening of
$q$ monopoles in the non oblique $SU(p)$ vacuum. We shall presently
show that the complete picture is more involved. 

We work in the convention where an element of $SL(2,\bf{Z})$, represented by
a matrix 
$
M = \bpm \alpha & \beta \\
                   \gamma & \delta
    \epm
$,
operates on the charge vector $(e , m)^t$ or the string vector 
$(F1 , D1)^t$ by multiplication, while on the 5--brane vector  
$(NS5 , D5)^t$ it operates by a multiplication by
$
\left(M^{-1}\right)^t = \bpm \delta  & -\beta \\
                                       -\gamma & \alpha
                        \epm
$.
We take the generator $T$ of $SL(2,\bf{Z})$ to be given by
$
T = \bpm 1 & 1 \\
         0 & 1
    \epm
$
and denote 
$
T' = S^{-1} T S =  \left(T^{-1}\right)^t = \bpm 1  & 0 \\
                                                -1 & 1
                                           \epm 
$.

Let us look at the $p$ D5 branes description of the non oblique  
$SU(p)$ vacuum. 
By performing a $(T')^b$ action, every D5 brane becomes a $(b , 1)$ 
5--brane, and the ``electric'' screening is now of $p \cdot (1 , b)$
charges. This amounts to changing to some oblique vacuum.
Let us now apply the same transformation to the $q$ NS5 branes
description of the non oblique $SU(p)$ vacuum. This has no effect on
$NS5$ branes, but the vacuum must transform in the same manner as
before!

The answer, of course, is that the ``magnetic'' screening is
changed. The $p \cdot (1 , 0)$ strings on the baryon vertex of the NS5
{\em are} affected by the action of $(T')^b$, 
and transform into $p \cdot (1 , b)$ 
strings, giving the same
vacuum as before. This example shows that the field--theoretic vacuum is 
{\em not} completely described by the 5--brane wrapping the $S^2$.
As $(p , 0)$ charges are screened ``electrically'' by the $p$ D5 wrapped
branes, the charges screened
``magnetically'' can be any of $(b , q)$, with the relevant values
of $b$ being $0 \le b < p$. These $p$ possibilities correspond to the
splitting of the classical $SU(p)$ vacuum into $p$ quantum vacua,
parameterized by the expectation values of the gaugino bilinear 
$\langle\lambda_4 \lambda_4\rangle$. 
It cannot be said that the $p$ D5 brane
configuration is canonically the non oblique $SU(p)$ vacuum -- all
the $SU(p)$ oblique vacua are on equal footing, and the condensation
choosing between them is another worldvolume phenomenon that cannot be
inferred from the SUGRA solution alone.
Note, however, that the action of $(T')^b$ also changes the NS
potential $B_{\mu \nu}$ ($G_3$ is invariant as $\tau$ is also changed).

A $p' \cdot (h , 0)$ 5--brane configuration representing $p'$ branes
(i.e having $\gcd(h , q') = 1$ with $p' \cdot q' = N$), will give
``magnetic'' screening  of $q' \cdot (1 , 0)$ charges.
This can be seen from the analysis of Polchinski and Strassler,
remembering that the baryon vertex, arising from the Hanany--Witten
effect, will consist of $h$ D3 branes filling the sphere, and those D3
branes can separate in the Minkowski space, each carrying $q'$
fundamental strings.
  
Generally, a $p' \cdot (c' , d')$ configuration can be obtained from a
$p' \cdot (h , 0)$  one using multiplication by 
\begin{equation}
\label{SLmatrix}
\bpm c'' & r'' \\
     d'' & s'' 
\epm                   \equiv
\bpm c'/h & r'/h \\
     d'/h & s'/h 
\epm                   \in SL(2,\bf{Z})
\end{equation}
and then the ``magnetic'' screening will be of $q' \cdot (s'' , r'')$
charges.
 
\subsection{Compatibility with the $\bf{Z}_N \times \bf{Z}_N$ 
Lattice Structure}

We have seen that a $(c , d) = p' \cdot (c' , d')$ 5--brane
configuration, with 
$q'$ and $h = \gcd(c' , d')$ mutually prime, has two generators of the
screened charges lattice -- the ``electric'' one 
$p' \cdot (d'' , c'') \equiv p' \cdot (d'/h , c'/h)$, 
and the ``magnetic'' one is some $q' \cdot (s'' , r'')$
where $p' \cdot q' = N$ and (\ref{SLmatrix}) holds. 

First we note that this solution gives screening
compatible with the $\bf{Z}_N \times \bf{Z}_N$ structure of the field theory
charges lattice. Indeed, from (\ref{SLmatrix}), the two vectors 
$(c'' , d'') \;,\; (r'' , s'')$ span the whole $\bf{Z} \times \bf{Z}$
lattice, so the two aforementioned generators certainly span $(N , 0)$
and $(0 , N)$. 
Therefore, we may concentrate on the $\bf{Z}_N \times \bf{Z}_N$ lattice  
spanned by the two generators modulu $N$. We summarize the configuration
in the matrix
\begin{equation}
\label{SLNmatrix}
\bpm c'' & d'' \\
     r'' & s'' 
\epm              \in SL(2,\bf{Z}_N)
\end{equation}
where we also attach to this matrix the number $p' \in \bf{Z}_N$.

Next, if there is some combination of the generators giving a trivial
charge, 
\begin{equation}
(a p' , b q')  
\bpm c'' & d'' \\
     r'' & s'' 
\epm               \equiv (0 , 0) \;\;\; \pmod N
\end{equation}
then by the invertibility of the matrix, $a \equiv 0 \;\pmod {q'}$ and 
$b \equiv 0 \;\pmod {p'}$.
The two generators therefore generate a lattice of size $p' \cdot q' = N$, 
corresponding indeed to a massive vacuum.

\newsec{A Classification of All Brane Configurations Leading 
to a Given Vacuum} 
\label{sec:classification}

Now that we have a full description of the vacua generated by wrapped
5--branes, we can try and classify all the configurations giving rise
to a given field theory vacuum. In this section we will restrict ourselves to
square--free $N$, that is $N$ which is a product of distinct
primes. For such an $N$, if $N = p' \cdot q'$, then $p'$ and $q'$ are
necessarily mutually prime.

Let us begin with an example involving the $SU(p)$ non oblique vacuum.
We know from \cite{PolStr} that it can be either represented by $p$ D5
branes or by $q$ NS5 branes. In the first description, the
generator $(p , 0)$ of the screened charges subgroup arises 
``electrically'' and the generator $(0 , q)$ arises ``magnetically'', 
while the roles are reversed in the second description.   
However, when $p$ and $q$ are mutually prime, the subgroup 
$P = \; \langle (p , 0) \,,\, (0 , q) \rangle$ is cyclic, 
and is generated by a single element: 
$P = \; \langle (p , q) \rangle$ (the choice of generator is, 
of course, non unique).
This is just the Chinese remainder
theorem, or the simple assertion that in these circumstances, 
$\bf{Z}_p \times \bf{Z}_q = \bf{Z}_{p q}$.
>From the previous discussions it follows that the same vacuum can also be
described by a single $(q , p)$ 5--brane, where the full screened
charges lattice is accounted for ``electrically'', and the ``magnetic''
screening accounts only for the modularity by $N$. 

In fact, when $N$ is square--free, the only Abelian subgroup of order
$N$ is $\bf{Z}_N$, and all the massive vacua subgroups of
the $\bf{Z}_N \times \bf{Z}_N$ lattice are isomorphic.   
When $p$ and $q$ are not mutually prime, the subgroup 
$P = \; \langle (p , 0) \,,\, (0 , q) \rangle$ is not cyclic, and therefore 
this does not hold for $N$ which is not square--free.

We know that the action of $T$ rotates between the different oblique
vacua corresponding to the same electric screening. From the non
oblique $SU(N)$ confining vacuum we can therefore get to all the
oblique analogues, where the ``magnetic'' screening can be $SU(q)$ for all 
$q | N$. Analogously, given $p | N$ the action of $T'$ can take us from the
confining vacuum to some vacuum with electric $SU(p)$, and then by $T$ to any
oblique or non oblique analogue of it. 
We see that for square--free $N$, the vacua are isomorphic by the
action of the $SL(2,\bf{Z}_N)$. Therefore it is enough to study
the representations of a single vacuum, e.g. the confining vacuum. 

When we have a $(c , d)$ brane, with $f = \gcd(c,d)$, 
we should first of all determine 
$p' = \gcd(f , N)$ and then 
$c'' = c / f, \; d'' = d / f$ (as a matter of fact, in
order to obtain $c'',d''$ it is sufficient, modulu $N$, to divide
$c,d$ only by the factor of $f$ consisting of all the primes
which divide $N$, to the appropriate power). These $c'',d''$ are
obviously given modulu $N$. Then, $r'',s''$ obeying (\ref{SLNmatrix})
should be found, but there are only $p'$ such pairs giving distinct
physical configurations. We can now ask the reverse question --
classify all the pairs $(c'',d'') \in \bf{Z}_N \times \bf{Z}_N$, 
mutually prime modulu $N$,
appropriate pairs $(r'',s'')$ and with an attached number $p' | N$,
giving a prescribed vacuum.

We find
that each vacuum is obtained exactly once from every $(c'',d'')$ pair,
with an appropriate $p'$ and $(r'',s'')$. 
The number of such representations for each vacuum is $\phi(N) \sigma_1(N)$, 
where $\phi(N)$ is Euler's totient
function, counting the number of $1 \le i \le N$ relatively prime to
$N$, and $\sigma_1(N)$ is the sum of the divisors of $N$. 
The details and examples are left for the appendix.
 
Not all the representations we have found are small perturbations of
the ${\cal N} = 4$ SUGRA dual solution. Indeed, for a shell of wrapped
$(c , d)$ 5--branes, with $M = c \tau + d$. 
The small curvature
condition is then $g |M| \gg 1$, and the small perturbation condition
\cite{PolStr} is $N \gg g |M|^2$. Using the
averages inequality, for the case without $\theta$ angle, we translate
the small perturbation condition to
\begin{equation}
N \gg g |M|^2 = 2 g \cdot \frac{1}{2} \left((\frac{c}{g})^2 + d^2\right) \ge 
2 g \cdot \left(\frac{c d}{g}\right) = 2 c d
\end{equation}
which is certainly satisfied
only for a very limited part inside the periodicity range considered
above. For example, the $(q , p)$ 5--brane solution of the $SU(p)$
vacuum is already (marginally) outside the range of validity.
However, even if the perturbations are not small, the first order
approximations to the SUGRA equations do not hold, and the explicit
solution of \cite{PolStr} is not valid, 
wrapped 5--branes still seem to provide the correct qualitative
behavior as described above.  

Be that as it may, our analysis shows, for example, that for odd $N$,
the $(0 , 2)$ 5--brane configuration yields the same Higgs vacuum as a
$(0 , 1)$ configuration (a single D5). Certainly, when the latter
description is valid, so is the former. Therefore, there inevitably are
multiple valid descriptions of the same vacuum. 
More generally, we may multiply a valid $(c,d)$ 5--brane configuration by a
small number relatively prime to $N$, to get another valid description
(in the appendix we also show that in a certain sense this is the
generic way to get multiple valid descriptions).

The parallel in field theory of the situation described in the last
paragraph is the non uniqueness of the generators of the 
condensed charges subgroups. In the field theory
classification of vacua, a generator has no meaning, only the group
does. In the the dual theory, however, different 5--brane
configuration may correspond to different generators, and moreover,
each generator has an infinity of 5--brane configurations giving it. 
This is the source of the multiple stringy descriptions of each field
theory vacuum.

All the above is contrary to the view of \cite{PolStr} that there is 
only one valid description for every choice of vacuum and value of
$g$.
Clearly this result is puzzling. If we believe in the AdS/CFT duality
and its generalizations, we expect to find only one valid SUGRA background
dual to a certain field theory (although, of course, when the
parameters of the field theory change, the valid description can also
change, as was shown in \cite{PolStr}). 
It is a challenging question to check whether this multi--description
is indeed present. There are several possible resolutions to 
this puzzle: 
\begin{enumerate}
\item
The different solutions may turn out to be in fact equivalent 
(therefore describing exactly the same field theory). This, however,
is unlikely, as it seems that the details of the physics probed by the
various strings (i.e. the exact potential between external charges), as
well as the physics detected by other probes (i.e. other branes and
bulk fields) would be different in the different solutions.
\item
The multiplicity present in
the level of the brane configuration analysis might be removed in the still
lacking full string theory treatment, i.e.
stringy effects beyond the SUGRA approximation with
brane dynamics might render some of the solutions invalid. 
The natural candidate solutions to be removed are those in which the
D3 charge is not divisible by the ``naive'' number of 5--branes.
This seems to us to be the most probable option.
\item
Some of the superfluous solutions might correspond, through some
stringy mechanism yet unknown to us (perhaps on the worldvolume of the
branes), to massless vacua. As there are many such vacua, and as they
are less well understood, this might help to solve the problem.
\item
Even within the SUGRA approximation some of the solutions may be non--BPS
and unstable. They would therefore represent valid non--vacuum field theory
states. These states would have non--zero energy density and
therefore would be ``false vacua''.  
\item
Following ideas \cite{GMT,HHI} similar to those associated with 
``giant gravitons'' \cite{McST}, 
it might be  possible that there are instantonic tunnelings between different
5--brane configurations giving the same field theory vacuum. In this case
the SUGRA solution should be taken as a superposition of these
configurations.
However, it
is not clear what is the field--theoretic analogue of such a
superposition. Moreover, there does not seem to exist in the field
theory a ``theta angle'' corresponding to superpositions with relative
phases.
\end{enumerate}

\newsec{Mass gap and asymptotic states}
\label{sec:glueball}

It was pointed out in \cite{PolStr} that $\N=1^*$ theory has a rich
spectrum of asymptotic states. Some of the expected states, like monopoles
and W-bosons, cannot be seen by the SUGRA solution presented in
\cite{PolStr}. These require a full analysis of the worldvolume
theory on the 5--branes referred to in section \ref{sec:screening}. 
The spectrum of states corresponding to bulk supergravity modes 
can be  calculated in much the same way as in \cite{witten1,oz,nunes}. 
One major difference is that $SO(6)$ non-singlet modes were not considered in
those papers, since they have no corresponding states in QCD$_3$. In
the $\N=1^*$ theory such states exist, and can potentially even be
lighter than the $0^{++}$ glueball which corresponds to the dilaton
 \cite{loewy}. 
Although the dual background presented
in \cite{PolStr} is not a product space with an $SO(6)$ isometry,
it becomes one near the boundary. We can thus still use the spectrum
derived in \cite{kim} to enumerate the SUGRA
modes and classify them according to their spin and other quantum
numbers. In principle, the non-trivial background fields can cause
mixing between SUGRA modes. In such a case the $0^{++}$ glueball does
not correspond simply to the dilaton, and one has to find
combinations of 
SUGRA modes that are eigenstates of the system of linearized type IIB
equations. 
We will not attempt to find such a combination here. 
Note, however, that because the background is asymptotically 
$AdS_5 \times S^5$, all possible mixings will be suppressed in the UV,
where the classical value of the dilaton is constant. 
Instead we will consider a fictitious minimally coupled scalar $\psi$ 
that has the following action in the Einstein frame:
\begin{equation} S_{\psi} \sim \int d^{10}x \ \sqrt{g_E} \  (\partial \psi )^2 \end{equation}  
In a generic vacuum state where $SU(N)$ is broken down to $SU(p)$ we
expect to find glueball  states, as well as W-bosons and
monopoles. Consider the background of p D5--branes.   The metric and
dilaton (\ref{metric}) can also be written as
\begin{equation} 
ds^2 = Z^{-1/2} dx_{0123}^2 + Z^{1/2} \Big[
du^2 + u^2 d \la^2 + u^2 \sin^2 \la d \Om_{y}^{2} \Big] +
\Om^{1/2}Z^{1/2} u^2 \cos^2 \la d \Om^2_w \end{equation}
\begin{equation} e^{2 \phi}=g^2 \Om^{1/2}, \ \ \ \ \ \ \ 
Z=\frac{R^4}{u^4-2r_0^2u^2 \cos 2 \la +r_0^4}, \ \ \ \ \ \ \  
r_0=m \ap \pi N /p, \ \ \ \ \ \ \ 
\Om = \Big[ \frac{\rho_-^2}{\rho_-^2 + \rho_c^2} \Big] ^2 
\end{equation}
where we introduced the coordinates $u^2=y^2+w^2$ and $\tan \la
=y/w$. Note that since both $y$ and $w$ are radii and therefore positive, $\la$
takes values in $[0,\pi/2]$.
 
Consider the wave equation for $\psi$,
\begin{equation} \partial_{\mu} 
\Big[ \sqrt{g_E} \  g_E^{\mu \nu}
\ \partial_{\nu} \Big] \psi = 0 \end{equation}
Explicitly, the equation is 
\begin{equation} 
\label{equ} 
\Big[ \D_u^2 + \frac{5}{u} \D_u + 
\frac{1}{u^2} ( \D_\la^2 + 4 \cot2 \la \ \D_\la ) - k^2 Z(u,\la) \Big]
\psi = 0 
\end{equation}
where $k^\mu$ is the 4-momentum along $x_{0123}$. The field theory asymptotic states associated with $\psi$ have  masses $M^2=-k^2$. 

In order to get a normalizable $\psi$ we need to impose 
boundary conditions at the boundary of the $(u,\la)$ plane 
\cite{witten1,oz,nunes}. The wave
function should obey $\psi(\infty, \la)=0$,  and there is an additional
singular point at $(r_0,0)$.  We should also specify boundary
conditions at $\la = 0$ and $\la = \pi/2$.
Since both $w$
and $y$ are radii of three dimensional spaces, 
smoothness of the wave function at
their origins implies $\D_w \psi(w=0,y)=\D_y
\psi(w,y=0)=0$. 
It is convenient to continue $\la$ to $[0,2 \pi]$ by reflection,
$y \ra -y$, $w \ra -w$, adding a singular point at
$(r_0, \pi)$ so that the wave function is automatically even with
respect to these reflections, and demand  $\psi(\infty, \la)=0$ 
for the enlarged range of $\la$.
We should also impose $\psi=0$ at the two singular
points in order for the solution to be normalizable.

We can easily show that $\psi$ has no normalizable zero mode. If one looks for
a solution with $k=0$ the equation (\ref{equ}) separates. The nontrivial
eigenvalues of the
operator $\D_\la^2 + 4 \cot2 \la \ \D_\la$ are $\alpha_n = -4(n^2 - 1)$ 
where $n \ge 1$  is an integer.
Therefore, equation (\ref{equ}) reduces to
\begin{equation} 
\Big[ \D_u^2 + \frac{5}{u} \D_u + \frac{\alpha_n}{u^2} \Big] 
\psi = 0 
\end{equation}
and it can be seen that there is no solution $\psi(u)$ that
obeys all boundary conditions. In order to prove the existence of a
mass gap one should in principle repeat such arguments for all the
possible SUGRA modes, after the wave equations of type IIB are diagonalized. 

The mass scale for the asymptotic states that one gets from
(\ref{equ}) is 
\begin{equation} 
M^2 \sim r_0^2/R^4 \sim \frac{m^2 N^2}{gNp^2 } = m^2 \frac{q}{gp} 
\end{equation}
To the extent that $\psi$ is a good approximation of one of the fields
in type IIB SUGRA, the above mass scale is that of the
glueball states corresponding to that field.
The ratio $gp/q$ is exactly the controlling parameter introduced in
\cite{PolStr}. For weak coupling (small $g$) where the $p$ D5 branes
background is valid the mass scale is therefore large. 
At strong coupling one should
use the $q$ NS5 branes background. The wave equation is the same, but
now $r_0=m \ap \pi g N /q$. This leads to the
following scale 
\begin{equation} M^2 \sim r_0^2/R^4 \sim \frac{m^2 g^2 N^2}{gNq^2 } = m^2
\frac{gp}{q} \end{equation}
Again, the ratio $gp/q$ is large when the $q$
NS5 branes background is valid. These results are consistent with, and
generalize, the analysis carried out in \cite{PolStr} for a single D5 and for
a single NS5 brane. As explained in \cite{PolStr}, in the $q$ NS5 branes
background there is a throat region, and it is not clear whether any
of the arguments above are valid.

\newsec{Global Symmetries and $\N=0^*$ Theory}
\label{sec:symmetry}

In this section we consider the route of supersymmetry breaking
from the $\N=4$ CFT, through $\N=1^*$, down to the non-supersymmetric
$\N=0^*$, concentrating on the global symmetries of the theories and
their manifestation in the string theory duals.
\begin{itemize} 

\item $\N=4$ SYM. \\
As mentioned above, the superpotential (\ref{supot})
manifestly exhibits a $U(3)$ R-symmetry. 
Writing the scalar fields as
\begin{equation}\label{Ai}
\phi_i = \frac{A_i + i A_{i+3}}{\sqrt{2}}
\end{equation}
the scalar potential is
\begin{equation}\label{V}
V \propto \sum_{i,j} tr \left( [A_i,A_j]^2 \right)
\end{equation} 
with a manifest $SO(6)$ symmetry. The symmetry is also visible in the dual
description since it is a part of the isometry of  the $AdS_5\times S^5$
metric. It is important to note that in this case there are no
additional background fields that are not invariant under this
symmetry.

\item $\N=1^*$. \\
In the field theory picture
the perturbation (\ref{pert}) breaks the R-symmetry to
$U(1)$  . This symmetry which transforms  the gauginos
is anomalous and  is  further broken by instantons to
  $\bf{Z}_{2N}$.
In \cite{MalNun} it was shown  in a different context that in the 
 SUGRA picture  worldsheet instantons break the $U(1)_R$ symmetry 
associated with a shift symmetry of one of the $S^3$ Euler angles. 
Due to  a flux of a $B$ field the  worldsheet 
instantons produce a phase that is proportional to $N$ times
the angle so only shifts of $\frac{2\pi n}{ N}$ are allowed, namely, a breaking
of the $U(1)_R$ down to $\bf{Z}_{2N}$. 
A  mechanism of a similar nature may also apply in our case. 

In the equal masses case, the scalar potential
\begin{equation}\label{Vper}
V \propto \sum_{i,j} tr \left([\phi_i,\phi_j][\phi_i,\phi_j]^\dagger
\right) + m^2 \sum_i| \phi_i |^2 + m \sum_{ijk} \epsilon^{ijk}\left(
[\phi_i,\phi_j]\phi_k^\dagger
+ [\phi_i,\phi_j]^\dagger \phi_k \right)
\end{equation}
exhibits an  $SO(3)$ global symmetry. 
However, as discussed in section \ref{sec:review}, this symmetry mixes
with the $SU(N)$ local gauge symmetry, and each vacuum possesses a new
$SO(3)$ symmetry.

It is presumably this new $SO(3)$ symmetry which   
is manifest as the isometry
of the SUGRA dual of the $\N=1^*$ theory, as we now proceed to show.  
Considering the metric
(\ref{metric}), the isometry group
seems to be $SO(3) \times SO(3)$. However,  this is not the whole picture.
 Unlike the $AdS_5\times S^5$ and some of its relatives, in the $N=1^*$
there is the additional  $G_3$ background field whose ``isometries'' have
to be examined separately. Since it is an important point, we show now
explicitly that in fact  $G_3$ is invariant only under the
diagonal subgroup $SO(3)_z$ of $SO(3)_w\times SO(3)_y$.   
The field $G_3$ obeys  
\begin{equation}
*_6 G_3 - i G_3 = Z'(w^2,y^2) T_3
\end{equation}
where $T_3$ is given by (\ref{T3}) with $m_1=m_2=m_3, m_4=0$. 
The metric and the function $Z'$ are invariant under
$SO(3)_w \times SO(3)_y$. 
The Hodge operator $*_6$ is defined relative to the {\em flat},
$SO(6)$ invariant metric.
In addition, the
Bianchi identity is, to lowest order, $d G_3 = 0$, and in the next
order of the perturbation there is a magnetic source, $d G_3 = J_4$,
with $J_4 = J_4(w^2,y^2)$ also invariant under $SO(3)_w \times SO(3)_y$.
The invariance of $G_3$ is therefore dictated by that of $T_3$. It can
easily be seen that $T_3$ is only invariant under the diagonal
subgroup $SO(3)_z$ of $SO(3)_w\times SO(3)_y$, in which 
$z_i = \frac{w_i + i y_i}{\sqrt{2}}$ and 
$\bar{z}^i = \frac{w^i - i y^i}{\sqrt{2}}$ transform in the
same manner. 
Explicitly, let us consider the orthogonal transformations which are
the exponentiations of (\ref{infiso3}),
keeping $w_3$ fixed, and taking, for any angle $\theta$,
\begin{equation}
\left( 
\begin{array}{c} w_1 \\ w_2 \end{array} 
\right)                                         \longmapsto
\left( 
\begin{array}{cc} \cos \theta & \sin \theta \\
                 -\sin \theta & \cos \theta
\end{array} 
\right)
\left(
\begin{array}{c} w_1 \\ w_2 \end{array}
\right)
\end{equation}
and similarly for $y_i$, therefore also for 
$z_i,\bar{z}_i$. 
We easily see that
both the two--form wedged with
$dz_3$ and the one wedged with $d\bar{z}_3$ are invariant, 
and therefore so is $T_3$. It can also be easily seen that there are
no transformations outside $SO(3)_z$ keeping $T_3$
invariant. Therefore, this should also be the invariance group of
$G_3$, and the global symmetry of the SUGRA solution.  
Alternatively, The explicit solution of $G_3$ given in \cite{PolStr} 
can be checked out to show the same invariance.
 
\item  $\N=0^*$, mass to the gaugino alone. \\
Supersymmetry may be completely broken by giving mass
to the gaugino. This can be done by perturbing the superpotential \cite{APSY}
\begin{equation}
\Delta W = M_4 S
\end{equation}
where $S$ is the glueball chiral superfield 
(including $tr(\lambda_4 \lambda_4)$) 
and $M_4$ is a chiral superfield whose F--term is $m_4$.
Obviously, such a perturbation breaks explicitly the $U(1)_R$ symmetry.
The $m_4$ component is  the part of the ${\bf \overline{10}}$ representation 
which is a singlet under the 
$SU(3)$ but carries a charge under the $U(1)_R$. 
If we assign a mass to the gaugino,
keeping all the other fields massless at tree level, the classical global
symmetry is $SU(3)$. This chiral symmetry is claimed to be broken by
the condensate $\langle \lambda_i \lambda_i \rangle$ of the 
chiral superfield fermions \cite{DisZam,ColWit}.
 
The corresponding SUGRA perturbation
is turned on  by the last term in (\ref{T3}). A mass  term only for the
gaugino
means that the terms proportional to $m$ are switched off.
This setup was shown in \cite{Zamo} to 
be outside the range of validity of the Polchinski--Strassler solution.
Once supersymmetry is broken, the scalars are no longer
protected and will inevitably acquire masses through radiative
corrections.

\item  $\N=0^*$, mass to the  scalars alone. \\
An alternative method to break supersymmetry while maintaining
the classical $U(1)_R$ symmetry is giving 
masses to the scalars.
The traceless part gives mass to  
$tr(\phi_i \phi_j - \frac{1}{6} \phi^2 \delta_{ij})$,
a chiral primary operator
transforming in the {\bf 20'} of $SO(6)$, which 
 corresponds \cite{kim,PorZaf} to the SUGRA field $S_2$ which is a
linear combination of the four--form and of the trace of the metric on $S^5$.
In the notation of \cite{PolStr} it is  $\mu_{mn}$.
We note that such a mass term gives rise to tachyonic modes. 
This may indicate that the theory does not possess a vacuum state. One
can contemplate that this problem may be cured non--perturbatively, by
generating a positive potential leading to a stable minimum,
corresponding to a true ground state. However, instantons do not seem to
create such a potential. Moreover, for large fields we expect the
semi--classical approximation to hold.
Another way this problem can be solved is by adding the trace. 
The trace part $tr(\phi_i \phi_i)$ is not a
chiral operator  and corresponds to a stringy mode.
Clearly we do not know how to implement this alternative or how to check it.
Note that this perturbation alone does not involve the $T_3$
and thus the creation of 5--branes via the Myers effect is absent here.
At the tree level there is an $SU(4)$ symmetry that protects the
masslessness of all the fermions.

\item $\N=0^*$, mass to the chiral superfields and to the gaugino. \\
Obviously, any perturbation caused by turning on either
$m_4$ or $\mu_{mn}$, combined with $m$, breaks supersymmetry completely.
The global symmetries associated with these models follow from those of 
the previous $N=0^*$ models. For instance,
the symmetry of the theory with $m$ and $m_4$ turned on is $SO(3)$. 
These theories were considered in \cite{Zamo} and the
string dual, practically identical to the one in \cite{PolStr}, indeed
possesses the $SO(3)$ symmetry.

\end{itemize}

\newsec{Summary and Discussion}
\label{sec:summary}

Supergravity duals of large $N$ gauge theories have provided a useful
laboratory to ``measure'' gauge invariant properties in the large
$\lambda$ regime. In particular, Wilson loops were computed in about a
dozen of SUGRA setups \cite{Cobi} and were shown to admit the behavior 
anticipated from gauge dynamics. 
Apparently, that is not the case for the dual of the $\N=1^*$ model.
A naive calculation of the minimal area of a string worldsheet 
in the metric of the
model yields ``wrong'' expectation values for the Wilson loops
(for instance, for the vacuum  associated
with the  $p$ D5 branes).        
The reason for the  failure of the naive approach in  the present case
is obviously twofold. First, the curvature of the metric diverges and
the SUGRA approximation ceases to hold. Moreover, there is 
an additional non-trivial background field, the magnetic
three--form. In fact the coupling of the string 
also to a $B^{(NS)}_{\mu\nu}$ occurs in SUGRA duals of certain field theories
on non-commutative spacetime \cite{MalRus, HashItz, Cobi2}. 
In those cases the coupling is well understood 
and the extraction of the Wilson loops indeed produced  
the anticipated  picture.
Unfortunately, at present the determination of the full string action 
in the $\N=1^*$ dual is beyond our abilities. 
However, as was shown in section 3,  the incorporation 
of the world volume gauge dynamics on the unwrapped part of the five-branes 
leads to a picture that at least qualitatively is in accordance  
with the properties of the field theory vacua. 

A natural question that arises is what should the string action look like  
had we known how to incorporate the coupling to the full background. 
Again let us 
address the case where the $SU(N)$ symmetry is broken down to $SU(p)$. 
The string tension
for such a case may have a factor of the form 
$\sin(\frac{ 2\pi k}{p})$ where $k$ is 
the $p$-ality of the external quarks. 
Such a  behavior was detected in two--dimensional 
systems \cite{AFS} as well as in MQCD \cite{HanStr}.
An interesting open question is  whether  a similar structure occurs also 
in the $\N=1^*$ dual. 
     
A challenging puzzle (that has been already mentioned in \cite{PolStr}) 
is the fact that there are multiple valid descriptions  for the same vacuum.
Contrary to \cite{PolStr}, 
it seems that in the SUGRA approximation, taking into account also
brane dynamics, there is 
more than one description corresponding to each vacuum and regime of
$g$.
We have explained why this result is perplexing, and have discussed
several possible resolutions of the problem.
This question certainly deserves further study. 

The  $\N=1^*$ theory serves as a bridge between  theories which are rich with 
supersymmetries and typically are non confining, and those with
${\cal N}=1$ or less that have confining phases.  
Recently, important steps toward the  construction
of a dual string model of  pure  ${\cal N}=1$ SYM theory 
were made \cite{KleStr,MalNun}. 
It would be interesting to compute Wilson--'t Hooft loops, 
analyze the vacuum structure, 
determine the glueballs spectrum etc. in these novel SUGRA backgrounds.     

\vspace{24pt}
{\large \bf Acknowledgments}

We thank Vadim Kaplunovsky for participation in the initial stages of
this work. J.S. would like to thank C. Gomez and 
E. Alvarez for numerous discussions while visiting their
institute. 
We would also like to thank O. Aharony for illuminating discussions.
This work was  supported in part 
by the US-Israel Binational Science
Foundation, by GIF -- the German-Israeli Foundation for Scientific Research,
and by the Israel Science Foundation.

\newsec{Appendix -- Counting of brane configurations}

First we prove that the number of pairs 
$(c'',d'') \in \bf{Z}_N \times \bf{Z}_N$ 
mutually prime modulu $N$ is 
$\phi(N) \sigma_1(N)$, where $\phi(N)$ is Euler's totient
function, counting the number of $1 \le i \le N$ relatively prime to
$N$, and $\sigma_1(N)$ is the sum of the divisors of $N$. 
The reason is as follows. Denote $v = \gcd(c'' , N)$, so that $v$ can
be any divisor of $N$, and write $c'' = v w$. $w$ is mutually prime to
$N/v$, so the number of its possible values, which is also the number
of possible $c''$, is $\phi(N/v)$. Now, $d''$ must be mutually prime
with $v$, so the number of its possible values is $\frac{N}{v} \phi(v)$.
$N$ is square--free, so $v$ and $N/v$ are mutually prime, therefore by a
fundamental property of Euler's function, the number of pairs $(c'',d'')$
for a given $v$ is 
$\phi(N/v) \cdot \frac{N}{v} \phi(v) = \frac{N}{v} \phi(N)$.
Finally, the number of $(c'',d'')$ pairs for any $v$ is 
$\sum_{v|N} \frac{N}{v} \phi(N) = \phi(N) \sum_{v'|N} v' = 
 \phi(N) \sigma_1(N)$.

Further we should select
$p'$ and appropriate $(r'',s'')$ for a given $(c'',d'')$ pair.
As there are in this case exactly $p'$ physically distinct $(r'',s'')$ pairs, 
this gives another factor of 
$\sum_{p'|N} p' = \sigma_1(N)$, so the total number of vacua 
representations is 
$\phi(N) \sigma_1(N)^2$. The number of distinct quantum massive vacua is 
$\sigma_1(N)$, as was explained in section \ref{sec:review}, 
and every vacuum is represented the same number of
times, as argued in section \ref{sec:classification}. 
Therefore, the number of representations for each vacuum is 
$\phi(N) \sigma_1(N)$, exactly as the number
of $(c'',d'')$ pairs. 

Indeed, we find that each vacuum is obtained exactly once 
from every $(c'',d'')$ pair,
with an appropriate $p', (r'',s'')$. 
This is easily seen for the confining vacuum, having screened charges 
subgroup $P = \langle (0 , 1) \rangle$.
For this vacuum, $q' \equiv \gcd(c'' , N)$ and $\gcd(r'' , N) = p'$,
so there are no screened particles with non zero electric charge modulu $N$.
As also explained in section \ref{sec:classification}, this result
can, by the action of $SL(2,\bf{Z}_N)$, be applied also to all the
other massive vacua for the square--free $N$ case.    

The distribution of representations regarding $p'$
is also egalitarian, each vacuum obtained $p' \phi(N)$ times from each
$p'$.

Elementary number theory shows that for square--free $N$, 
$\phi(N) = \prod_{p|N} (p-1)$ and $\sigma_1(N) = \prod_{p|N} (p+1)$.
We shall now give an example for the smallest non--trivial case, $N=6$, 
for which the number of representations for each vacuum is already
$\phi(6) \sigma_1(6) = 2 \cdot 12 = 24$. We choose to display results
for the non--oblique $SU(3)$ vacuum, for which the subgroup of
screened charges is 
$P = \left\{(0,0) , (0,2) , (0,4) , (3,0) , (3,2) , (3,4)\right\}$.
We repeat that the choice of $(r'',s'')$ for a given $(c'',d')$ 
is not unique, as $(r'',s'')$ and 
$(r'' + p' c'' , s'' + p' d'') \; \pmod N$ yield the same generator
(after multiplication by $q'$). However, the physical outcome is
identical for two (or more) such $(r'',s'')$ pairs.


The first row of table 1 
corresponds to 
$(c,d) = p' \cdot (c'',d'') = 3 \cdot (0,1) = (0,3)$, or to the three D5
brane configuration of \cite{PolStr}. 
The third row corresponds to  
$(c,d) = p' \cdot (c'',d'') = 2 \cdot (1,0) = (2,0)$, or to the two NS5
brane configuration of \cite{PolStr}, while the tenth row is the
$(2,3)$ 5--brane solution introduced in section \ref{sec:classification}.

Now we move to make exact the claim of section \ref{sec:classification}
that two valid 5--brane descriptions of the same vacuum are small
multiples of the same basic configuration. 
We assume that the two descriptions have the same $p'$. In fact, we
shall take $p'=1$ in the following, as the general case is very similar.
We pick some $p$ (with $p \cdot q = N$) and demand that 
both descriptions involve less than 
$q$ NS5 branes and less than $p$ D5 branes. This is natural for valid
descriptions, for we know \cite{PolStr} that $p$ D5 branes 
and $q$ NS5 branes cannot be both valid descriptions for the same value of $g$.
Let the two 5--branes be $(c_1,d_1)$ and $(c_2,d_2)$. The counting
argument given above shows that 
$
(c_2,d_2) \equiv n \cdot (c_1,d_1) \; \pmod N
$
for some
$n$ mutually prime with $N$, but a priori the magnitude of $n$ is
unknown to us. Multiplying the first element of this equation by 
$d_1$, the second by $d_2$, and subtracting, we get
$c_2 d_1 - c_1 d_2 \equiv 0 \; \pmod N$.
But as by assumption, $0 \le c_1 , c_2 < q$ and $0 \le d_1 , d_2 < p$,
we must have 
$c_2 d_1 - c_1 d_2 = 0$.
Therefore $c_2/c_1 = d_2/d_1$, and both solutions are indeed
(small) multiplications of some basic one.   


\begin{table}[h]
\begin{center}
$
\begin{array}{cc}
\begin{array}{l|c|c|c}
   & (c'' , d'') & (r'' , s'') & p' \\
\hline
 1 & (0,1)       & (5,0)       & 3 \\
 2 & (0,5)       & (1,0)       & 3 \\
 3 & (1,0)       & (0,1)       & 2 \\
 4 & (1,1)       & (2,3)       & 6 \\
 5 & (1,2)       & (4,3)       & 6 \\
 6 & (1,3)       & (0,1)       & 2 \\
 7 & (1,4)       & (2,3)       & 6 \\
 8 & (1,5)       & (4,3)       & 6 \\
 9 & (2,1)       & (5,0)       & 3 \\
10 & (2,3)       & (5,5)       & 1 \\
11 & (2,5)       & (1,3)       & 3 \\
12 & (3,1)       & (2,3)       & 6 
\end{array}
 \hspace{24pt} & \hspace{24pt}  
\begin{array}{l|c|c|c}
   & (c'' , d'') & (r'' , s'') & p' \\
\hline
13 & (3,2)       & (4,3)       & 6 \\
14 & (3,4)       & (2,3)       & 6 \\
15 & (3,5)       & (4,3)       & 6 \\
16 & (4,1)       & (5,0)       & 3 \\
17 & (4,3)       & (1,1)       & 1 \\
18 & (4,5)       & (1,3)       & 3 \\
19 & (5,0)       & (0,5)       & 2 \\
20 & (5,1)       & (2,3)       & 6 \\
21 & (5,2)       & (4,3)       & 6 \\
22 & (5,3)       & (4,5)       & 2 \\
23 & (5,4)       & (2,3)       & 6 \\
24 & (5,5)       & (4,3)       & 6 
\end{array}
\end{array}
$
\end{center}
\caption{List of all configurations giving rise to the non oblique
  $SU(3)$ vacuum of $SU(6)$ $\N=1^*$ theory}
\end{table}


\end{document}